\RequirePackage[l2tabu, orthodox]{nag}
\documentclass{article}
\pdfoutput=1
\usepackage[preprint,nonatbib]{neurips_2020}
\usepackage[sort&compress,numbers]{natbib}
\usepackage[
    colorlinks=true,citecolor=carmine,linkcolor=carmine,urlcolor=googleblue
    ]{hyperref}
\usepackage{amssymb}
\usepackage{amsmath}
\usepackage{diagbox}
\usepackage{graphicx}
\usepackage{multirow}
\usepackage[utf8]{inputenc}

\usepackage[dvipsnames]{xcolor}
  \definecolor{purple}{RGB}{128,0,128}
  \definecolor{carmine}{RGB}{150,0,24}
  \definecolor{googleblue}{RGB}{34, 0, 204}
\usepackage{fontawesome}
  \newcommand{\at}{{\scriptsize\faAt}}

\author{Aidan Kehoe \\
GraphStax Research \\
\texttt{akehoe\at graphstax.com}
\AND
Peter Wittek \\
Rotman School of Management - University of Toronto \\
Creative Destruction Lab \\
Vector Institute for Artificial Intelligence \\
Perimeter Institute for Theoretical Physics
\AND
Yanbo Xue \\
Career Science Lab - BOSS Zhipin \\
\texttt{xueyanbo\at kanzhun.com}
\AND
Alejandro Pozas-Kerstjens \\
Department of Mathematical Analysis - Universidad Complutense de Madrid \\
\texttt{physics\at alexpozas.com}
}

\begin{document}
\title{Defence against adversarial attacks using classical and quantum-enhanced Boltzmann machines}

\maketitle

\begin{abstract}
We provide a robust defence to adversarial attacks on discriminative algorithms.
Neural networks are naturally vulnerable to small, tailored perturbations in the input data that lead to wrong predictions.
On the contrary, generative models attempt to learn the distribution underlying a dataset, making them inherently more robust to small perturbations.
We use Boltzmann machines for discrimination purposes as attack-resistant classifiers, and compare them against standard state-of-the-art adversarial defences.
We find improvements ranging from 5\% to 72\% against attacks with Boltzmann machines on the MNIST dataset.
We furthermore complement the training with quantum-enhanced sampling from the D-Wave 2000Q annealer, finding results comparable with classical techniques and with marginal improvements in some cases.
These results underline the relevance of probabilistic methods in constructing neural networks and highlight a novel scenario of practical relevance where quantum computers, even with limited hardware capabilites, could provide advantages over classical computers.

This work is dedicated to the memory of Peter Wittek.
\end{abstract}

\section{Introduction}
The robustness of machine learning models against adversarial attacks is currently an open question. Indeed, it is easy to make neural networks misclassify images by applying perturbations to just one single pixel~\cite{su2017singlepixel}, or even perturbations that are imperceptible to human observers. Such a perturbation is found by maximizing the network's prediction error~\cite{szegedy2013intriguing,goodfellow2014explaining}, therefore showing adversarial examples which rely on the inherent uncertainty that neural networks have about their predictions~\cite{cubuk2017intriguing}. These ideas lead to real-world attempts at fooling state-of-the-art computer vision models~\cite{moosavi-dezfooli2016deepfool} and enabling malware to bypass detection~\cite{srndic2014practical,grosse2017adversarial}.

Right now there exist a great plethora of adversarial attacks on neural-network discriminative algorithms.
The most common are white-box, meaning that they have access to the neural network parameters after training~\cite{szegedy2013intriguing,goodfellow2014explaining,cisse2017houdini,carlini2016towards}.
In contrast, black-box attacks just query the classifier to construct adversarial examples~\cite{brendel2018decision,zhao2018generating}.
These latter attacks are more difficult to create but potentially more efficient to a wide range of classifiers, not just deep neural networks.
However, it has been shown that substitute models can be trained on black-box queries to a target model, and adversarial examples built from white-box attacks to the substitute model are also effective in the target~\cite{papernot2016transfer}.
It is hotly debated whether there is a type of machine learning model, or any data processing technique, that is most adept at defending against adversarial examples.
Indeed, right now defences are often broken soon after their publication~\cite{carlini2017adversarialnoteasy,carlini2017magnet,madry2018towards,fawzi2018adversarial}, and part of the community's proposal is encouraging a faster iteration cycle between attacks and defences via competitions~\cite{kurakin2018adversarial}.

Bayesian methods and generative probabilistic graphical models (PGMs) could potentially be more robust, since their training is not necessarily related to backpropagation and first-order gradient descent.
Gaussian processes, for instance, achieved a remarkable performance against attacks~\cite{bradshaw2017adversarial,grosse2017how}. 
In a similar way, generative models, which learn the full joint probability distribution of instances and labels instead of an input-output function, are thought to be more resistant to attacks based on small perturbations.
The fact that the full data manifold is approximated in generative models was the inspiration behind generative adversarial networks (GANs)~\cite{goodfellow2014generative}, that were originally designed such that a generator would approximate the data manifold to fool a discriminator.
Our motivation stems from this observation and we ask the question whether generative probabilistic models can have an improved robustness to adversarial attacks when compared against feedforward neural networks.

In this work we benchmark the robustness of generative PGMs against recent state-of-the-art attacks on discriminative neural networks, and assess the performance of state-of-the-art defences on these attacks.
We compare the results previously known for feedforward deep classifiers against classical and quantum-enhanced restricted Boltzmann machines in white-box attack schemes.
Our findings show that both classical and quantum-enhanced Boltzmann machines far outperform the current competition, with improvements ranging from 4.62\% to 72.41\%.
Moreover, quantum-enhanced Boltzmann machines give marginal improvements over their classical counterparts in some instances, which is an encouraging result given the young state of quantum hardware platforms, and illustrates a practical application of interest of noisy, intermediate-scale quantum technologies
more direct than
demonstrations of quantum superiority~\cite{superiorityGoogle,superiorityChina}.

This paper is organized as follows:
in Section~\ref{sec:adversarial} we introduce the concept of adversarial attacks and review the relevant recent work in this area.
We proceed analogously in Section~\ref{sec:freeenergy} for the training of restricted Boltzmann machines.
Then, in Section~\ref{sec:experiments} we describe the particular architectures, attacks and defences we consider, and evaluate them in Section~\ref{sec:results}.
We conclude in Section~\ref{sec:conclusions} with a discussion and pointing to directions of future work.

\section{Adversarial examples}\label{sec:adversarial}
Formally, an adversarial attack is aimed at a classifier $f: \mathcal{X}\mapsto \mathcal{Y}$, where $\mathcal{X}$ is typically a subspace of $\mathbb{R}^d$, and $\mathcal{Y}$ is a discrete space corresponding to labels.
The goal of the attack is to create an example $\mathbf{x}^*$ in the vicinity of a valid training point $\mathbf{x}$ such that $\|\mathbf{x}^*-\mathbf{x}\|$ is small and $f(\mathbf{x}^*)\neq f(\mathbf{x})$.
If $f(\mathbf{x}^*)$ does not have any restriction other than $f(\mathbf{x}^*)\neq f(\mathbf{x})$ , the attack is said to be non-targeted. In contrast, in a targeted attack the adversary aims to specify the label of the adversarial example, that is, $f(\mathbf{x}^*)=y^*$ for a specific choice of $y^*$.
The difference to the actual examples should be small as measured by the $L_p$ norm, where $p$ is typically 0, 1, 2, or $\infty$.
For a more complete taxonomy, see Refs.~\cite{brendel2018decision,kurakin2018adversarial,yuan2017adversarial}.

As an illustration of a standard attack, let us consider the fast gradient sign method (FGSM)~\cite{goodfellow2014explaining}. This attack is interesting since it approximates the minimization in the infinity norm bound, \mbox{$\|\mathbf{x}^* - \mathbf{x}\|_{\infty}$}, in a single step, which means that it can scale easily to large datasets.
FGSM is an example of a white-box attack, where the model parameters are known to the adversary.
Adversarial examples are created by

\begin{equation}
  \mathbf{x}^* = \mathbf{x} + \epsilon\cdot\mathrm{sign}(\nabla_{\mathbf{x}}\mathcal{L}_\theta(\mathbf{x},y)),
  \label{eq:fgsm}
\end{equation}

where $\nabla_{\mathbf{x}}\mathcal{L}_\theta(\mathbf{x},y)$ is the gradient of the loss function $\mathcal{L}_\theta$ of the machine learning model parametrized by $\theta$ in $\mathbf{x}$, whose true label is $y$.
In its simplest formulation, FGSM does not transfer well to other models than the one it was trained on because of the fact that all the information needed to prepare the adversarial examples is very specific to the model used.
On the other hand, black-box schemes in turn either transfer attacks from an undefended model to other models or query an unknown model to gain information about its decision boundary.
In both cases, the agnosticity about the target model make black-box attacks more difficult to craft but potentially more ``dangerous'' in terms of their reach.

Adversarial examples in linear models can be explained as a result of high-dimensional dot products.
In such models,  the output when a perturbed image is input will be the sum of the dot product of the weights with the unperturbed input and the dot product of the weights with the perturbation.
If each weight vector has average magnitude $m$ and the input is $n$-dimensional, the activation in the classifier will grow by $\epsilon m n$ (where $\epsilon$ is the magnitude of the perturbation).
For low-dimensional problems this does not result in a large difference in classification between the original and perturbed data, as long as the activation does not exceed the precision of the features.
However, if the input has a large dimension $n$, even weak perturbations can cause large errors in the classifier~\cite{goodfellow2014explaining}.

There are other hypotheses that suggest that adversarial examples also arise in neural networks despite their non-linear nature, but there is very little convincing evidence for them.
In fact, many commonly used neural network architectures such as long short-term memory networks~\cite{hochreiter1997long}, maxout networks~\cite{goodfellow2013maxout}, and rectified linear units~\cite{jarrett2009what} intentionally behave linearly in order to reduce computational complexity, increasing its vulnerability to adversarial examples.
However, this does not preclude non-linear architectures of neural networks such as sigmoid networks to suffer from similar problems.
Many of these types of networks work mostly within a linear regime to better optimize their results, thus also leaving them open to adversarial examples as well.

While the theoretical underpinning of deep neural networks is improving, the robustness guarantees of the defence mechanisms were primarily supported by empirical observations, occasionally masking implicit assumptions.
In turn, this leads to an arms race of devising defences, which are very soon bypassed by more elaborate attacks~\cite{carlini2017adversarialnoteasy}.

Certain models are naturally immune to adversarial examples: kernel methods, for instance, provide an example of models with this type of immunity~\cite{goodfellow2014explaining,hein2017formal}.
Immunity does not come for free, since kernel methods are not competitive at all with deep neural networks in terms of performance on clean datasets.
This seems to suggest that performance and robustness against attacks are two important features of machine learning algorithms that are at odds, and that when developing an application based on neural networks one must choose between making it perform well or making it secure.
This idea was formally expressed and proved in Ref.~\cite{Papernot2016}, where it is shown that robustness to adversarial attacks can be achieved by employing models of higher complexity, provided that there are no bounds on the access to new datapoints so the model underlying the data can be learned with high fidelity.
The general case of a learning scheme is, however, markedly different, as the learner is typically provided with a dataset of fixed size.
In the case of fixed datasets, the use of more complex models leads to overfitting, which is itself a source of weakness to adversarial attacks, and models with high capacity that fail to be as robust as models with a lower capacity are necessarily overfitting.

Methods other than neural networks may provide an advantage in this case. Bayesian inference methods have been shown to identify deviations in the uncertainty about predictions between adversarial and clean examples. This was demonstrated with Gaussian processes~\cite{bradshaw2017adversarial,grosse2017how}, and in fact, statistical tests alone can already indicate whether samples were drawn from the same distribution as the original data~\cite{grosse2017on}.
Furthermore, Gaussian processes have a direct correspondence between deep and wide neural networks~\cite{lee2018deep}, allowing for a potentially high accuracy on clean examples, addressing the main shortcoming of kernel methods.

It still remains unclear whether other PGMs, and in particular deep variants, are more robust against adversarial examples.
Many of these models are quintessentially generative, including the well-known family of Boltzmann machines.
In backpropagated architectures, generative models are not necessarily robust against adversarial perturbations~\cite{goodfellow2014explaining}, but generative probabilistic models are yet to be benchmarked.

\section{Minimizing free energy}\label{sec:freeenergy}
While PGMs can still use some variant of gradient descent for training, the error is not backpropagated in the sense of feedforward neural networks.
In fact, training involves a so-called negative phase (also known as sleep cycle or thermal equilibration) that is global in the sense that it possibly factors in everything the network has previously seen, adding some stochastic variations, and it also involves global, long-range connections.
Therefore, we rightfully expect that PGMs are robust against attacks enabled by the kind of local manipulations associated to backpropagation.

Looking at it from a different angle, backpropagation with stochastic gradient descent aims at obtaining a good local optimum, whereas the minimization of free energy, which is the goal when training energy-based PGMs, gives an average over all local optima and the global optima.
While an attack can easily nudge backpropagation to a different local optima, the free energy already considers all local optima, and hence its expected robustness towards perturbations.

In this work we focus on a particular type of PGMs, known as restricted Boltzmann machines (RBMs).
The training of RBMs is computationally efficient with heuristics such as contrastive divergence (CD) and persistent contrastive divergence (PCD).
Formally, the function to minimize when training RBMs is the negative log-likelihood,

\begin{equation}
  \mathcal{L}_\theta\left(\mathcal{T}\right)=-\frac{1}{|\mathcal{T}|}\sum_{\mathbf{x}^{(i)}\in\mathcal{T}}\log\left(p_\theta(\mathbf{x}^{(i)})\right),
  \label{loss}
\end{equation}

where $\mathcal{T}$ is the training set and $\theta$ are the parameters of the model.
Intuitively, one intends to maximize the product of the probabilities of the instances in the training set, which should appear frequently (i.e. with higher probability) since they represent ``good'' configurations of the visible units.
In fact, the ideal function to minimize is the Kullback-Leibler (KL) divergence between the probability distribution that the RBM learns, $p_\theta(\mathbf{x})$, and the real probability distribution over the instances $p(\mathbf{x})$.
Discarding the term that does not depend of the model parameters from the KL divergence and assuming an approximately equal probability for good configurations to appear, Eq.~\eqref{loss} is recovered.

The parameter update rule is computed by calculating the derivatives of Eq.~\eqref{loss} with respect to $\theta$.
For simplicity, let us choose a Boltzmann distribution in a system with hidden neurons, namely $p_\theta(\mathbf{x})=Z_\theta^{-1}\sum_\mathbf{h}\exp[-E_\theta(\mathbf{x},\mathbf{h})]$ where the sum over all possible configurations of the units defines the partition function, \mbox{$Z_\theta=\sum_{\mathbf{x},\mathbf{h}}\exp[-E_\theta(\mathbf{x},\mathbf{h})]$}.
Moreover, let us define the free energy of a configuration of visible units $\mathcal{F}_\theta(\mathbf{x})$ by

\begin{equation}
e^{-\mathcal{F}_\theta(\mathbf{x})}=\sum_\mathbf{h} e^{-E_\theta(\mathbf{x},\mathbf{h})}.
\end{equation}
Using this, the loss function can be rewritten as
\begin{align}
  \mathcal{L}_\theta\left(\mathcal{T}\right)=\log \sum_\mathbf{x} e^{-\mathcal{F}_\theta(\mathbf{x})} + \frac{1}{|\mathcal{T}|}\sum_{\mathbf{x}^{(i)}\in\mathcal{T}}\mathcal{F}_\theta(\mathbf{x}^{(i)}).
  \label{eq:loss_fe}
\end{align}
The second sum is over the training set, while the first one is over \emph{all} possible neuron configurations.
This is the term whose computation is intractable classically---for a model with $N$ binary neurons, the total number of possible configurations is $2^N$.
Now, computing the gradient of Eq.~\eqref{eq:loss_fe}, we observe two different terms:
\begin{align}
  \frac{\partial\mathcal{L}_\theta}{\partial\theta}(\mathcal{T}) =  \frac{1}{|\mathcal{T}|}\sum_{\mathbf{x}^{(i)}\in\mathcal{T}}\frac{\partial\mathcal{F}_\theta(\mathbf{x}^{(i)})}{\partial\theta}  -\sum_\mathbf{x}\left[p_\theta(\mathbf{x})\frac{\partial\mathcal{F}_\theta(\mathbf{x})}{\partial\theta}\right] .
\label{lossgrad}
\end{align}
Here we identify the \textit{positive} phase as the term coming from the evaluations on the training set, and the \textit{negative} phase as that coming from evaluations on all possible configurations of visible units.
The simplest energy function associated to an RBM is given by

\begin{equation}
  E_\theta(\mathbf{x},\mathbf{h})=-\sum_i x_i b_i - \sum_j h_j c_j -\sum_{ij} w_{ij} x_i h_j,
  \label{eq:energyrbm}
\end{equation}
where $\theta$ is now $\{w_{ij}, b_i, c_j\}_{ij}$. The free energy can be thus expressed as

\begin{equation}
  \mathcal{F}_\theta(\mathbf{x}) = -\sum_i x_i b_i -\sum_j\log\left(1+e^{c_j+\sum_i w_{ij} x_i}\right).
  \label{freeenergy}
\end{equation}

Picking any derivative, we can see a more explicit formulation of the positive and negative phases.
For instance, if we take the visible biases, $b_i$, with Eq.~\eqref{lossgrad} we obtain
\begin{align}
  \frac{\partial\mathcal{L}}{\partial b_i}(\mathcal{T})&=\frac{1}{|\mathcal{T}|}\sum_{\mathbf{x}^{(k)}\in\mathcal{T}}(-x_i^{(k)})  -\sum_x p_\theta(\mathbf{x})(-x_i) \notag \\
  & =-(\langle x_i\rangle_\text{data}  -\langle x_i\rangle_\text{model}).
\end{align}
The positive phase, $\langle\cdot\rangle_\text{data}$, is easy to compute since it only involves averages over the training set, but the negative phase, $\langle\cdot\rangle_\text{model}$, is not possible to calculate explicitly since it requires knowledge of the overall distribution $p_\theta(\mathbf{x})$, or analogously of the partition function $Z_\theta$.

\subsection{Classical heuristics}
To evaluate the gradient correctly one needs to be able to satisfactorily approximate the negative phase, $\langle\cdot\rangle_\text{model}$.
A way of obtaining samples from the model distribution is via Markov chain Monte Carlo Gibbs sampling~\cite{murphy2012machine}: starting from a random configuration of the visible neurons, one iterates sampling the hidden and visible neurons from the corresponding conditional distributions until the stationary state is reached.
This proper estimation is prohibitively expensive due to the large amount of iterations required to reach the stationary state, but some simple heuristics work decently in practice.

CD is one of such simple heuristics~\cite{sutskever2010convergence}.
Instead of iterating infinitely over the layers of the RBM to find $\langle\cdot\rangle_\text{model}$, when using CD-$k$ one performs $k$ iterations of a batch of $n$ chains that are initialized in training datapoints, and computes the expectation as averages over those $n$ chains.
This is an extremely simplified way of doing Gibbs sampling without a burn-in time, and instead of starting from random initialization, one starts from a clamped set of visible nodes.
Choosing to start from training data in the negative phase has two benefits: (i) the resultant gradient will produce a model that does not drift too far away from the ground truth, and (ii) there is no need to wait for the Markov chain to mix in order to draw samples of reasonable quality.

A more reliable variant of CD is called persistent CD (PCD)~\cite{tieleman2008training}, where instead of initializing the chains for every average to be computed,
the initial seeds for a sampling process are the final state of the chains in the previous process.
This may reduce the impact of burn-in since a Markov chain is constantly maintained.
However, one has to carefully balance between the mixing rate and weight updates~\cite{tieleman2009using}.

While seemingly useful in practice, one must bear in mind that by using these heuristics one may lose visibility of discovering samples that follow more closely the Boltzmann distribution.
Moreover, these sampling heuristics obviate that, especially in the early stages of training a Boltzmann machine, the fact that the weights are initialized randomly makes the proper characterization of the low-energy sector an NP-complete problem~\cite{Barahona1982complexity,SpinGlassControl}.

\subsection{Quantum-enhanced sampling}\label{sec:qes}
A powerful realization that has arisen with the recent advent of analog quantum computers is that many algorithms which had an inspiration in the dynamics of physical processes and that simulated or approximated those physical processes can now be readily implemented in analog computers, without approximations.
In the context of the NP-complete problem of computing the negative phase of a Boltzmann machine, many state-of-the-art methods have inspiration from physics, such as simulated annealing~\cite{SimulatedAnnealing} or parallel tempering~\cite{ParallelTempering}.
Now it is possible not just to simulate these dynamics, but to prepare a physical system in the particular desired state and measure or sample it directly in order to compute $\langle\cdot\rangle_\text{model}$.
It is well known that more accurate approximations of the negative phase give rise to improvements in terms of model quality and training time~\cite{tieleman2008training,SpinGlassControl}.
Thus, it is reasonable to expect that the direct access to model samples from measuring a physical system results in higher-quality models that, from the perspective of adversarial attacks, are more robust than models that are trained using classical heuristics.
Alternatively, one can reasonably expect to match the robustness of models trained using classical heuristics in fewer training cycles by using analog sampling methods.

In the early training stage of a Boltzmann machine, the Markov chains used in many heuristics for approximating the log-likelihood gradient may behave as expected because the weights are usually initialized as small values, which means that the Markov chains just need to approximate samples drawn from distributions close to thermal noise.
At later stages, however, it becomes harder to keep the chains near their stationary distributions~\cite{salak2012}.
It is now possible, however, to prepare systems in the stationary distribution directly, instead of approximating it.
The D-Wave quantum annealer is a physical device that has a working mechanism similar to simulated annealing~\cite{lanting2014entanglement}.
In addition to using the thermal fluctuations inherent to any physical system, quantum annealing (QA) further explores the search space by exploiting quantum fluctuations.
Studies have shown that large-scale quantum annealers such as those exemplified by the D-Wave quantum computers can offer significant speedups for certain problem classes~\cite{denchev2016what}.

Ref.~\cite{dmytro2016benchmarking} conducted extensive experiments that compare the sampling performance of QA against CD.
For energy functions with high barriers QA is more effective in exploring complex search spaces, which leads to a more accurate calculation of gradients.
Computational efficiency aside, CD and PCD may not yield high quality models that are competitive with backpropagated architectures in terms of prediction accuracy, despite the fact that they might be more robust to attacks.
This has inspired further heuristics such as Boltzmann-Encoded Adversarial Machines~\cite{fisher2018boltzmann}, where a GAN-like setting is combined with an RBM architecture.
Nevertheless, CD and PCD are just approximations of thermal sampling training schemes, which can be performed exactly by hardware-based sampling to yield intrinsically high-quality Boltzmann machines.
In particular, quantum-enhanced sampling on contemporary quantum annealers has already shown notable advantages against CD and PCD training~\cite{benedetti2017quantum,benedetti2017quantumhelmholtz}.

When working with QA one must make a distinction between two different distributions.
The first one is the actual Boltzmann distribution with energy function defined in Eq.~\eqref{eq:energyrbm}.
The second one is the quantum annealer sampling distribution. In theory, once the binary units are replaced by qubits as in QA, the energy function~\eqref{eq:energyrbm} is substituted by the Hamiltonian~\cite{amin2016quantum}
\begin{equation}
  H_\theta(\sigma^z_x, \sigma^z_h) = \sum_x b_x\sigma^z_x + \sum_h c_x\sigma^z_h + \sum_{x,h}w_{xh}\sigma^z_x\sigma^z_h,
  \label{eqn:qbm}
\end{equation}
where $\sigma^z_i$ represents a spin on a lattice site $i$  that takes a value in $\{+1, -1\}$.
Technically, these spins can be in a superposition, which means that they are equivalent to qubits.
    Eq.~\eqref{eq:energyrbm} will represent the diagonal elements of the $2^N{\times}2^N$ Hamiltonian matrix described by Eq.~\eqref{eqn:qbm}, and $N$ is the total amount of neurons of the Boltzmann machine.
The partition function then becomes the trace of $e^{-H_\theta}$, i.e., $Z_\theta = \mathrm{Tr}[e^{-H_\theta}]$.

Unlike classical Boltzmann machines where approximations have to be made to circumvent the evaluation of the partition function, in a quantum annealer one can prepare physical samples quantum mechanically, without the need of any explicit calculation. 
Instead, the samples drawn from the annealer directly follow the desired Boltzmann distribution.

\begin{figure}
	\centering
	\includegraphics[width=0.4\textwidth]{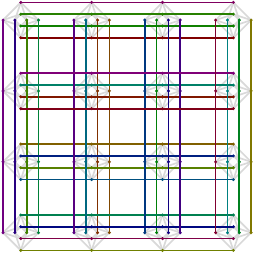}
	\caption{Embedding of a $16{\times}16$ RBM on a D-Wave Chimera graph of $4{\times}4$ unit cells. This illustrates the sparse connectivity of the hardware architecture and explains how multiple physical qubits form a logical binary variable: the coloured edges connecting the physical qubits represent a logical binary variable. The number of physical qubits need for a logical one is called chain length. The longer the chain, the higher the chance of a failed calculation. The grey couplings are not used in the embedding.}
    \label{fig:chimera_c4}
\end{figure}

Despite its sampling capacity, qubits on the D-Wave quantum computer are sparsely connected, which means that not any two qubits can be connected (or coupled) by a $w$.
In the current generation of the D-Wave chip, namely D-Wave 2000Q, each qubit is connected to a maximum of six qubits.
In order to couple qubits that are far apart one needs to resort to embeddings.
An embedding requires building virtual qubits by chaining up multiple physical qubits with strong ferromagnetic couplings. Shorter chains are favoured against longer ones, since in general, the longer the chain, the more likely it will break, which would distort the result of the sampling.
In Fig.~\ref{fig:chimera_c4} we illustrate the embedding of a $16{\times}16$ RBM on a $4{\times}4$ D-Wave Chimera graph, where each unit in the RBM is composed of four connected qubits on the quantum chip, as shown by the coloured chains.
In this work, we use the heuristic {\em minor-embedding} method to map the graphical structure of the RBM architectures employed to the D-Wave Chimera structure~\cite{cai2014practical}.

Another important issue one needs to address  when performing QA is the so-called {\em temperature effect}.
There are two factors to the temperature effect.
First, the programmable parameters on the D-Wave chip are normally constrained within a specific range.
When an RBM is being trained, one needs to scale the model to make sure it fits within the parameter range on the chip.
Scaling the energy function of Eq.~\eqref{eq:energyrbm} can be considered as changing the temperature of the problem.
Second, the chip runs at a very low temperature---around the milli-Kelvin level---which implies that when the Hamiltonian of Eq.~\eqref{eqn:qbm} is solved on the D-Wave quantum annealer, it is actually solved with a scaling factor of $T\,{\ll}\,1$.
The combination of these two factors will ultimately determine the temperature at which the samples are returned.

\section{Experimental Setup}\label{sec:experiments}
We outline now the different procedures for attacking and defending classifier algorithms, and detail the parameter choices for the classical and quantum RBMs.

The general attack-defence procedure in neural networks is as follows: first, each of the defences is applied to the original dataset, producing ``defended'' training and test datasets.
Then, each ``defended'' training dataset is employed to finetune a ResNet18 convolutional neural network (CNN), and each ``defended'' test set is used to build datasets of adversarial images using each of the attacks.
The attack datasets are 1:1 with the test datasets, so every image in each defended dataset has an adversarial version in each of the attack datasets.
Finally, the robustness to adversarial attacks is measured as the percentage of test adversarial images that are misclassified by the CNN.

The choice of attacks and defences were chosen to reflect methods that are commonly used and that proved strong in previous benchmarks~\cite{kurakin2018adversarial}.
While the defences make use of no knowledge about the classifier they will defend (i.e., they are black-box), we consider white-box attacks where the adversary has access to the classifier's architecture and parameters.
Black-box attacks, which only have access to queries of the model, are in principle more ``dangerous'' in that they can easily apply to a wide variety of classifiers.
Prior work has proven, however, that it is possible for an attacker to train a model on black-box queries of the target, and use this white-box substitute model to build attacks that are effective in the target model~\cite{papernot2016transfer}.
Thus, considering just white-box attacks is realistic and sufficient.
Finally, all the attacks are non-targeted: the focus is put in creating a misclassification, rather than forcing a particular label to be predicted.

As for the proposal of this work (recall, replacing the CNN classifier by an RBM generative model that is later repurposed into a classifier), the defence consists of simply training the RBM on the original dataset. The set of adversarial attacks are built from the original dataset as well.
\subsection{Attacks}
\textbf{Fast Gradient Sign Method}
FGSM is a standard attack for generating adversarial images. The method has been described in detail in the Related Work section.

\textbf{Carlini \& Wagner}
Carlini and Wagner proposed gradient-based attacks that have been found to be among the most effective when using small perturbations~\cite{carlini2016towards}. It re-frames generating adversarial examples as an optimization problem. Clearly, this is a difficult problem to solve and multiple techniques are employed to simplify the optimization. Firstly, a binary search algorithm is employed to find a suitable coefficient for the optimization terms, and after that, the optimization terms are converted to \textit{arctanh} space so that efficient, state-of-the-art optimization solvers can be utilized.

\textbf{DeepFool}
This method uses a different insight to derive efficient attacks~\cite{moosavi-dezfooli2016deepfool}.
It characterizes robustness as the distance between a data point and the decision surface.
The aim of the adversary is to minimize the perturbation in the $L_2$ norm while misleading the classifier. DeepFool represents the faces of a polyhedron with the decision boundary planes from the classifier to describe the output space. The attack then finds the minimal perturbation that changes the classifier's decision within the polyhedron. 

\subsection{Baseline defences}
\textbf{Adversarial Training}
The most popular defence is plain adversarial training, in the spirit of GANs~\cite{szegedy2013intriguing}.
In contrast with the use of GANs for generative purposes, in the case of adversarial training it is the discriminator, and not the generator, that which one is interested in improving its performance.
This method injects adversarial examples from the attacks during the training phase, so the discriminator is trained either on a mix of both clean and adversarial examples, or on only the latter.
However, since the adversarial examples are generated with a specific type of attack, the discriminator remains vulnerable to other types of attacks.

Adversarial effects can also be efficiently mitigated through randomization, effectively ending up to a form of data augmentation~\cite{xie2018mitigating}.
Techniques commonly used include random resizing and random padding.
The rationale behind these defences is that iterative attacks could overfit specific network parameters, hence low-level image transformations can destroy the structure of adversarial perturbations.

This defence is of a nature different than the remaining, and thus is implemented in a different way.
Instead of applying the defence to the training dataset and training the CNN in it, in adversarial training the CNN is trained directly in the union of the original clean dataset and the result of the application of the attack to the original clean dataset.

\textbf{Feature Squeezing} 
Feature squeezing~\cite{weilinxu18featsqueeze} is a computationally inexpensive yet powerful state-of-the-art method for defending against adversarial attacks.
This method reduces the colour bit depth of each pixel in the images used for training.
When representing images, colour bit depths are often employed to display images that are very close to their natural counterpart.
However, the features that are created as a result of this colour bit depth are often not necessary for recognizing what an image is representing.
Therefore, reducing the colour bit depth can in theory also reduce the opportunities for an attack to find an adversarial image without sacrificing classifier accuracy.

\textbf{Spatial Smoothing}
Local spatial smoothing~\cite{weilinxu18featsqueeze}, which is related to feature squeezing, is a method consisting of reducing image noise.
It can be thought of as a sliding window that centers around each pixel in the image and replaces the pixel with the median value of each of the neighbouring pixels.
This creates a ``blur'' over the image and helps mitigate the effects of adversarial perturbations (especially when applied to salt-and-pepper noise), while maintaining the features that make correct classification possible.

\subsection{Sampling-based defences}
The method we propose, and furthermore we show effective for defending against adversarial attacks in the next section, is abandoning the concept of training a discriminative neural network and instead training a generative model on the joint dataset of inputs and corresponding targets.
This is, instead of parametrizing a function $f_\theta(x)$ by a neural network architecture and using backpropagation to minimize some notion of distance between the prediction, $f_\theta(x)$, and the corresponding true label, $y$, we propose learning the joint distribution of datapoints and corresponding labels, $p_\theta(x,y)$---provided as the Boltzmann distribution of some parametrized energy function---, by optimizing the negative log-likelihood given in Eq.~\eqref{loss} where the input $\mathbf{x}$ is given as the concatenation of a datapoint $x$ and its corresponding label $y$.
Once the model is trained, the label assigned to a datapoint is that which minimizes the free energy of Eq.~\eqref{freeenergy} when the neurons encoding the datapoint are fixed.
In the case of a small possible choice of labels, this can be determined by direct computation of the free energy of all the neuron configurations $\mathbf{x}_\ell=(x,\ell)\,\,\forall\ell=1,\dots,|\mathcal{Y}|$ that correspond to the concatenation of the datapoint and a valid label, and choosing the label for which the free energy is lowest, this is,
\begin{equation}
    y(x) = \underset{\ell}{\text{argmin}}\,\mathcal{F}(\mathbf{x}_\ell)~\text{s.t.}~\mathbf{x}_\ell=(x,\ell).
\end{equation}
In more complicated cases, one can resort to sampling the label from the conditional distribution $p_\theta(y|x)$ via Markov Chain Monte Carlo methods.

In the particular case of this work we focus on RBM models, where the computation of the positive phase of the gradient update is easy.
We implement them using the \textit{ebm-torch} package~\cite{ebm-torch}, setting the number of hidden neurons to 40, the learning rate to 0.01, and using the Adam algorithm for updating the parameters in the graph.
No dropout or other regularization methods are used.

We use two different techniques for estimating the negative phase.
On one hand we do a classical training, using PCD to obtain the approximations of $\langle\cdot\rangle_\text{model}$.
We choose a small batch size of 10 and long iteration chains ($k\,{=}\,50$).
On the other, we compute it directly from the physical sampling of a thermal quantum state of a set of qubits in the D-Wave 2000Q processor.
In this case we use a larger batch size of $1\,000$ for computing the positive phases, which is a more natural choice for the hardware.
In light of the temperature effect discussed in Sec.~\ref{sec:qes}, we rescale the coefficients manually to have a tight control over the temperature at which the annealing is done.
We request $500$ samples from the chip, which are classically postprocessed by the D-Wave server-side software stack.
Each quantum sample is obtained with a $20\mu s$ annealing time.
To mitigate the intrinsic control errors that might exist on the quantum chip, we randomly choose 5 spin reversal transformations, which means that the Hamiltonian remains unchanged if, along with the random flipping of spins, the signs of bias and coupling terms are flipped accordingly.  
Since we can quickly reach a reasonably good model with heuristic methods, we bootstrap the quantum-enhanced RBMs with a partially trained classical RBM pre-trained over 100 epochs and leave the heavy lifting to the quantum annealer at a later training stage.

\section{Evaluation}\label{sec:results}
MNIST provides a simple and robust baseline that is useful to understand some limitations of defence mechanisms~\cite{tramer2018ensemble}.
While arguably it should not be the only benchmark~\cite{carlini2017adversarialnoteasy}, we rely on it exclusively because this is the only commonly used image dataset that can easily be transformed to a current quantum processing unit. 
A larger image resolution or additional color channels would require an extensive transformation pipeline or the addition of convolutional layers before the RBM, which could introduce confounding effects in the evaluation.

Tables~\ref{tbl:28by28} and \ref{tbl:7by7} summarize the results.
We used two versions of MNIST: the original one, containing $28{\times}28$ greyscale images, and a rescaled one, containing $7{\times}7$ binary images.
The latter transformation was necessary to embed the RBM in the quantum annealer, as the original version would require a significantly higher number of qubits and connectivity than what is available with contemporary technology.

The columns of the tables refer to the defences.
The RBM is the classical version trained with PCD, whereas the QRBM in Table~\ref{tbl:7by7} is the version trained using the D-Wave quantum annealer.
The attacks are all implemented using the \textit{foolbox} library~\cite{foolbox}.

\begin{table}[ht]
  \centering
  \begin{tabular}{l|c|c|c|c}
    & Adv. Training & Feat. Squeezing & Spatial Smoothing & RBM  \\ \hline\hline
    FGSM & {13.39\%} & {20.24\%} & {18.54\%} & \textbf{81.18\%}\\ \hline
    DeepFool & 12.80\% & 19.10\% & 19.70\% & \textbf{83.36\%} \\ \hline
    CarliniWagner& 72.00\% & 78.54\% & 45.80\% & \textbf{83.16\%}
  \end{tabular}
  \caption{Accuracy of defences after attacks on the original MNIST test dataset ($10\,000$ $28{\times}28$ greyscale images of handwritten digits).
  The values in bold highlight the best-performing defence for each attack.
  }
  \label{tbl:28by28}
\end{table}

\begin{table}[ht]
  \centering
  \begin{tabular}{l|c|c|c|c|c}
    & Adv. Training & Feat. Squeezing & Spatial Smoothing & RBM & QRBM \\ \hline \hline
    FGSM & {19.36\%} & {28.03\%} & {14.74\%} & \textbf{87.15\%} & 84.49\% \\ \hline
    DeepFool & 21.03\% & 26.64\% & 13.66\% & 84.50\% & \textbf{85.18\%} \\ \hline
    CarliniWagner & 48.24\% & 77.15\% & 22.00\% & \textbf{85.71\%} & 82.92\%
  \end{tabular}
  \caption{Accuracy of defences after attacks on the MNIST test dataset downscaled to $7{\times}7$ and binarized. This transformation is necessary for training the RBM on the quantum annealer.
  The values in bold highlight the best-performing defence for each attack.
  }
  \label{tbl:7by7}
\end{table}

RBMs prove significantly more robust against attacks even when classical sampling heuristics are used during training.
This underlies the importance of architectures that have an internal equilibration step, as opposed to purely greedy backpropagation of errors.
This benefit of using the RBM can be tracked back to the generative nature of the model.
Instead of learning a ``simple'' class-conditional density like discriminative models, the RBM seeks to approximate the entire data distribution.
While this is a more difficult problem that may result in a lower performance on an unperturbed dataset, it also reduces the overfitting of the model, making it more capable of generalizing to adversarial images.
Referring back to Ref.~\cite{Papernot2016}, this result is an extension of the upper bound of the loss function for more complex models.
Clearly, complex models drawn from a rich class of hypotheses must have loss functions smaller than simpler models, given (and this is an important point) that enough training data is provided.
While the CNNs on which the baseline defences were established are much more complex than an RBM, their performance against adversarial examples demonstrates that they overfit to the training and test data.

Importantly, quantum-enhanced sampling is on par with the classical RBM results and beats backpropagation-based architectures.
While it only outperforms classical PCD in only one attack and by a small margin, given the current immature state of quantum technologies, this is already a remarkable result.
Furthermore, it highlights an important aspect of the use of quantum technologies in machine learning which is more directly applicable and easier to measure than the speedup claims based in computational complexity arguments~\cite{zhao2019bayesian}.

\section{Conclusions and discussion}\label{sec:conclusions}


We have shown that using generative PGMs as discriminators produces models that are much more robust to adversarial attacks than standard discriminators based on deep neural networks.
RBMs trained on the MNIST dataset are notably more robust to misclassification of crafted images designed to fool the classification than standard CNNs, even when defensive pre-processings of the datasets are applied in order to protect the training pipeline.
Moreover, using quantum-enhanced estimations of the log-likelihood gradient of the RBM resulted in comparable improvements, having one instance, the DeepFool attack, where the quantum-enhanced RBM was more robust than its classical counterpart.

The source of the robustness of generative PGMs to adversarial attacks can be ascribed to the overfitting to training (and available test) data that discriminative models present when trained on finite datasets.
However, given that the primary goal of the generative model is not the classification accuracy but the efficient approximation of the data manifold, its utility is lower than the achievable with discriminative models.
This is an expression of the no-free-lunch results of Ref.~\cite{Papernot2016} applied to training on finite datasets.

Further support of the claim that the non-discriminative learning procedure of PGMs is a major factor in the robustness of the model can be found in Ref.~\cite{MLcritique}.
It is observed that human learning relies mostly on a highly structured learning mechanism and unsupervised learning rather than supervised.
Since humans are much better than current machine learning techniques at defending against adversarial examples, mimicking how we learn may be the best way to design robust defences.
Moreover, graphical models in general are structured in a much more nuanced way than neural networks.
The structure of most graphical models encodes some prior knowledge about the nature of the system it is modelling, something that seems to be also present in the human brain.
The similarities of graphical models to the way humans learn not only points to a reason why RBMs may be more robust to adversarial examples than neural networks, but also suggests that other, more structured graphical models may be even more effective.

From the quantum perspective, quantum computers are often touted for a significant speedup expressed in terms of computational complexity or runtime in some specific setting.
However, the current mismatch between theoretical requirements for provable speedups and experimental feasibility motivates the exploration of alternative metrics by which quantum computing may provide advantages~\cite{weber2020optimal}.
Here we shift the attention from a speedup to a qualitative difference that quantum computers could offer in training energy-based PGMs.
It is known that the quality of the trained model increases and the number of training epochs decreases with better approximations of the negative phase of the gradient (a simple example is shown in Ref.~\cite{SpinGlassControl}).
In the context of robustness to adversarial attacks, it is thus expected that the improvements in training that direct access to samples drawn from the model distribution (via directly measuring the qubits in the D-Wave quantum annealer) enables give rise to more robust models using fewer computational resources.
The results that we present do not yet allow to make a strong claim regarding an advantage, but they signal to a promising research direction where quantum advantages may be easier to achieve in the current era of noisy, intermediate-scale quantum computers.

	
Although we are encouraged by the robustness of Boltzmann machines trained with QA to adversarial attacks demonstrated at this stage (which is, recall, comparable to those of models trained using classical heuristics), the size of the available quantum chips is not yet scalable to industry-standard datasets.
This is not only due to the number of qubits, but also to their sparse connectivity.
In fact, the sparse connectivity requires that, currently, most models of interest must be embedded in the chip's architecture by using additional qubits.
Available embedding heuristics are designed to preserve the ground state, but access to the low-energy excited states is also needed when training Boltzmann machines.
An important direction of subsequent research, that does not necessarily require experimental advances, is thus understanding the relations between the energy spectra of the original problem and its embedding in the chip.

In future work, we plan to extend the current analysis beyond RBMs.
By using a richer class of models---like fully connected Boltzmann machines---one can expect that its larger expressibility allows to improve the performance of classification while maintaining the adversarial robustness.
However, even simple modifications like lateral connections being introduced into the hidden layer require hardware sampling---or alternative approximations---to be used also when computing the positive phase. 
Even if one can extend the range of Boltzmann machines that can be trained, the number of qubits needs a dramatic increase for performing an end-to-end training in the quantum processing units.
As Ref.~\cite{benedetti2017quantumhelmholtz} pointed out, classical preprocessing is recommended.
A natural way to extend this work is to include convolutional layers and attach an RBM to the final layer. 

We considered just white-box attacks since these can also be built when one only has black-box access to the target model~\cite{papernot2016transfer}.
The application of the attacks to classical RBMs is straightforward.
For the quantum RBM, in contrast, white-box access (i.e., the access not to the parameters that are input to the annealer, but those characterizing the Boltzmann distribution of the samples) requires extensive calls to the quantum annealer, which is prohibitive at the moment, particularly if we factor in the tuning requirements of both the hardware and the Boltzmann machines.
Nevertheless, as the software stack improves and the cost of access drops, we expect that the white-box scenario will eventually become viable.

Other than the D-Wave quantum annealer, there are currently further choices of analog sampling platforms.
For example, degenerate optical parametric oscillators can simulate an Ising model, with the desired belief that they do not suffer from sparse connectivity~\cite{qnncloud2017qnn}.
Furthermore, fully classical application-specific integrated circuits are being tested for digital annealing~\cite{fujitsu2017digitalannealer}.
It is interesting to see if the robustness achieved with the D-Wave quantum annealer applies to other hardware technologies, which would indicate that the improvements are indeed consequence of the use of PGMs in classification, rather than of some other feature of quantum annealing.

\section*{Acknowledgments}
This work was done before Peter Wittek left us, and is dedicated to his memory. Peter was a great leader and mentor, which had great impact not only in the authors' careers, but in that of so many others. Peter, you would have enjoyed so much the advances we are witnessing. The QML community still misses you, and will keep doing so for a long, long time.

A.P.-K. is grateful to the Creative Destruction Lab for their hospitality. A.P.-K.'s work was supported by Fundaci\'o Obra Social ``la Caixa'' (LCF/BQ/ES15/10360001) and the European Union's Horizon 2020 research and innovation programme - grant agreement No 648913.
\bibliographystyle{apsrev-custom}
\bibliography{adversarial_rbm_bibliography}

\end{document}